\documentclass[aps,prl, reprint,superscriptaddress]{revtex4-1}
\usepackage{graphicx}
\usepackage{amsmath}

\begin{document}


\title{Focusing Waves at Arbitrary Locations in a Ray-Chaotic Enclosure Using Time-Reversed Synthetic Sonas}


\author{Bo Xiao}
\affiliation{Electrical and Computer Engineering Department, University of Maryland, College Park, Maryland 20742-3285, USA}
\author{Thomas M. Antonsen}
\author{Edward Ott}
\author{Steven M. Anlage}
\affiliation{Electrical and Computer Engineering Department, University of Maryland, College Park, Maryland 20742-3285, USA}
\affiliation{Physics Department, University of Maryland, College Park, Maryland 20742-4111, USA}

\date{\today}

\begin{abstract}
Time reversal methods are widely used to achieve wave focusing in acoustics and electromagnetics. Past time reversal experiments typically require that a transmitter be initially present at the target focusing point, which limits the application of this technique. In this paper, we propose a method to focus waves at an arbitary location inside a complex enclosure using a numerically calculated wave excitation signal. We use a semi-classical ray algorithm to calculate the signal that would be received at a transceiver port resulting from the injection of a short pulse at the desired target location.  The time-reversed version of this signal is then injected into the transceiver port and an approximate reconstruction of the short pulse is created at the target. The quaility of the pulse reconstruction is quantified in three different ways, and the values of these metrics are shown to be predicted by the statistics of the scattering-parameter $|S_{21}|^2$ between the transceiver and target points in the enclosure over the bandwidth of the pulse. We experimentally demonstrate the method using a flat microwave billiard and quantify the reconstruction quality as a function of enclosure loss, port coupling and other considerations.
\end{abstract}


\maketitle

\section {Introduction}
Wave focusing through a strongly scattering medium is an intriguing research topic in the fields of optics, acoustics and electromagnetics \cite{SLM0, Zhou2014, Ma2014}. Its potential applications include medical imaging, ultrasound therapy, communications, and nondestructive testing. In optics, wavefront-shaping has been used to spatially focus light both through and inside strongly scattering media \cite{SLM0,SLM1,SLM2}. One can also achieve focusing in the temporal domain using a time-reversal mirror (TRM). The time reversal technique was first developed in acoustics \cite{fink06,fink01,fink02,fink03,fink04,fink05.1,fink05.2}. Much work has been done to study the underlying theory and possible applications in target identification, detection and imaging \cite{0266-5611-18-6-320, 1504963, TRM4, TRM6, Liou2010115,Mora201278, 6502757, 6553232}. A TRM can work both in open systems with a strongly scattering medium placed between the target and transceiver ports \cite{TRM2,TRM5}, or in closed reflecting walled systems (`billiards') supporting ballistic propagation of waves in which the wavelength is much smaller than the billiard size \cite{TRM1,Lerosey1,TRM3,Yeh1}. In fact, a relatively simple single-channel TRM can be efficiently implemented in ray-chaotic billiard systems \cite{TRM1}, and the experiments discussed here are performed in such billiards.

\par Previous time reversal experiments typically employ two steps \cite{TRM1,TRM2,Lerosey1,TRM3}. First, in the time-forward step, one injects a short pulse at the target port and collects the resulting long-duration transmitted signal (called the ``sona'') at the transceiver port. In the time backward step, one time-inverts the previously collected and recorded sona signal and sends it back into the system through the transceiver port, hopefully resulting in a time reversed short pulse at the target port. Since an active source must be present at the target location to create the initial signal, and because the sona is unique to that location, this process must be repeated for any location upon which one desires to focus waves. As shown in previous work \cite{matt2,matt1,sun1}, one can relax this constraint to some extent by placing a passive nonlinear object at the desired target location and using its higher harmonic nonlinear response as a unique ``beacon'' for later time-reversal. In acoustics, several methods \cite{acoustic1, acoustic2} have been developed to shift the location of the reconstruction, but these are either limited to small shifts ($10\%$ range shift of the focal spot) or to the special geometric case of acoustic waveguides. In both of these cases one must still have a source located at a representitive target location to produce a baseline sona signal.

\par One concern about the time-reversal process is the reliability of a time-reversed sona signal to create a reconstruction as the scattering environment evolves and changes over time. For example the reconstruction quality of electromagnetic waves in a three-dimensional billiard was shown to be quite sensitive to the dielectric constant of the gas filling the enclosure \cite{bini1}. In fact, this extreme sensitivity of the reconstruction to details in the scattering environment has been exploited as a new sensor technology \cite{bini2,bini3}. In this paper we wish to create robust reconstructions at arbitrary locations that are less sensitive to details. This is one of the motivations to rely on the presence of stable geometrical properties of the billiard that give rise to robust``short orbits" that connect the wave-entry and wave-focusing points \cite{bini2, Yeh1}.

\par  Here we present a \textit{synthetic sona} method for focusing electromagnetic waves at an arbitrary location in a ray-chaotic billiard using an extension of the time reversal technique. We choose a ray-chaotic system because its ergodicity ensures that all rays launched into the system will visit all points on the billiard boundary. It is also the most challenging situation for our wave focusing technique because small errors in the initial ray trajectory will accumulate exponentially in time. Our method is successful, but has limitations due to wave propagation loss, port coupling mismatch, finite mode density of the billiard, and the existence of chaos in the ray limit. We discuss the effects of these factors by presenting experimental results on both high-loss and low-loss billiards, different antennas and frequency ranges (to modify coupling), and modifications of the cavity that vary the boundaries and modal structure. In general, we find that the synthetic sona method can produce good time-reversal focusing at an arbitrary location in lossy ray-chaotic billiard experiments with well-coupled antennas. 

\par The synthetic sona method requires numerically calculating the sona collected at a receiving port that is generated by a source at the target port. Here we utilize semiclassical methods to do this. Compared with other numerical methods, such as finite-difference time-domain (FDTD) computation of billiard scattering properties, the semiclassical method is more efficient when the wavelength is much smaller than the system size. When going to smaller wavelengths, most numerical methods require a finer grid which significantly increases the computational cost. In contrast, the semiclassical method has the same computation complexity in all frequency ranges.

\par In the following, we first describe our experimental setup and procedures, including the calculation of a synthetic sona signal, performing a time-reversal experiment in the time domain and also in the frequency domain. Then we introduce several metrics to measure the reconstruction quality, and we discuss factors that limit the reconstruction quality, such as loss and mismatched port coupling.

\section{Experiment}
\subsection{Calculation of Synthetic Sona}
\begin{figure}
\includegraphics[width=0.35\textwidth]{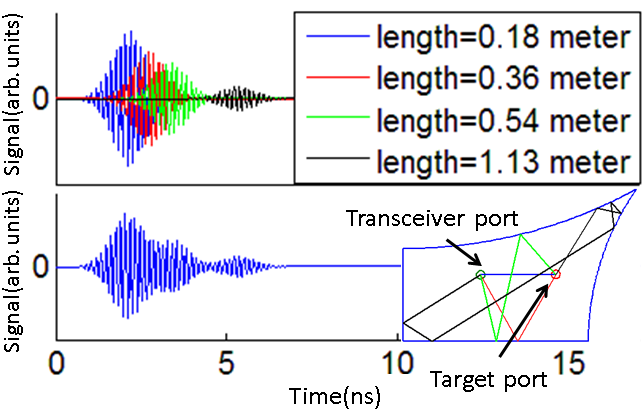}
\caption{\label{fig:fig1}Calculation of a brief synthetic sona from four simple orbits in a representitive 2D 1/4-bow-tie billiard. We first calculate a scaled and time-delayed version of the input signal, which is a Gaussian pulse modulation of a 7 GHz carrier signal in this example. These waveforms are summed to obtain the synthetic sona shown at the bottom. The two ports are 17.5 cm apart.}
\end{figure}

The construction of the synthetic sona starts with a calculation of ray orbits \cite{Yeh1} in the billiard. Specifically, limiting consideration to ray paths below a specified length limit, a ray tracing code is used to obtain the trajectories of rays that start from the target point, bounce off of the walls, propagate ballistically between bounces, and arrive at the transceiver port. Each bounce on the billiard wall follows the law of specular reflection, and we do not consider scattering from the ports. Then, for each trajectory $i$, the orbit length $L_i$, number of bounces $n_i$, and ray bundle divergence factor $D_i$ \cite{james1, Yeh3, Yeh1}, are used to calculate a scaled and time-delayed version of the input signal, $g(t)$, which is usually a short (on the order of the typical ballistic propagation time between bounces in the billiard) Gaussian pulse. Summing up contributions from all $N$ trajectories of length less than the upper limit gives the synthetic sona signal $s_{\text{syn}}(t)$. In practice the calculation is performed in the frequency domain first, $S(\omega)=\sum_{i=1}^{N}{G(\omega)e^{-j\omega L_i/c}(-1)^{n_i}\sqrt{D_i}}$, where $S(\omega)$ and $G(\omega)$ are the Fourier transforms of $s_{\text{syn}}(t)$ and $g(t)$ respectively. Then an inverse Fourier transform of $S(\omega)$ into the time domain gives  $s_{\text{syn}}(t)$. Figure \ref{fig:fig1} shows an example of a calculated synthetic sona from four simple orbits linking the target port and transceiver port in a 2D billiard. The above calculation does not include propagation loss. If we assume that the loss is uniform and results in an amplitude decay of $e^{-t/\tau}$ with amplitude decay time $\tau$, and also assume that $\tau$ is approximately frequency-independent, then we can apply an exponential window function to the synthetic sona to simulate the effect of propagation loss \cite{bini4}. 

\subsection{Experimental Setup}
For our microwave time reversal experiments \cite{TRM3, TRMour2}, the billiard is a quasi-2D, ray-chaotic cavity. That is, it is thin in one dimension (z) so that, at the frequencies of interest, the modes of the cavity have electric fields $\mathbf{E}=E_z(x,y) \hat{z}$ \cite{9780511524622}. One of the cavity shapes that we employ is depicted in the lower right inset of Figure 1 and is refered to as a symmetry-reduced `bowtie' shape \cite{Paulso,Dongho,sameer1}. We also utilize a superconducting Pb-coated cut-circle shape \cite{cutcircle2,cutcircle1,yeh2,yeh5,draeger1999one, draeger1999one-exp} at 7.01 K to create a billiard with minimal loss. 

\par
\begin{figure}
\includegraphics[width=0.45\textwidth]{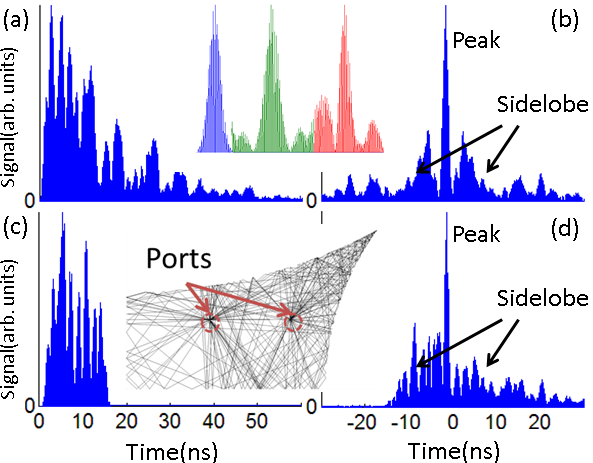}
\caption{\label{fig:fig2} (a) Physically measured sona, and (c) synthetic sona signal calculated from orbits less than 4 m ($=10\sqrt{A}$) in length, and their time reversal reconstruction signals, (b) and (d), respectively. Only the upper half of the signals are plotted since they are essentially symmetric about the time-axis. The upper inset shows closeups of the initial Gaussian pulse (left, blue), the measured sona reconstruction (middle, green) and the synthetic sona reconstruction (right, red). The bottom inset plots some of the orbits used to calculate the synthetic sona. The horizontal and vertical straight walls of the billiard have lengths of 43.18 cm and 21.59 cm, and the two ports are 17.5 cm apart. }
\end{figure}

 For comparison, we also employ a method based on the technique used in previously published time-reversal experiments. We generate a short Gaussian modulation pulse of a given carrier frequency, $g(t)$, inject it into the billiard through the target port, and a signal $s(t)$, called the \textit{sona} signal, is measured at the transceiver port (see Fig.\ref{fig:fig1} inset and Fig.\ref{fig:fig2}(a)). This sona signal is recorded and then time reversed. The time reversed waveform is then regenerated as a signal which is sent back into the billiard through the transceiver port. The signal $r(t)$ is then measured at the target port and, as desired, is found to approximately reconstruct the original Gaussian short pulse. The antennas used for the broadcast and receiving port have two-dimensionally isotropic radiation patterns, and are short metal pins extending from the center conductor of the end of coaxial transmission lines at port holes in the upper plate of the two-dimensional cavity \cite{sameer1}. The antenna has a 3dB bandwidth from 6.7 GHz to 11.9 GHz. Due to the variation in eigenmode amplitude at the transceiver and target points \cite{draeger1999one, draeger1999one-exp} and propagation loss, $r(t)$ not only contains a time reversed Gaussian pulse replica, but also has temporal sidelobes (see Figure 2b) which are symmetric about the reconstruction, to good approximation.

\par The envelope of $g(t)$ is a Gaussian of width $\sigma_t\approx0.5$ ns modulating a carrier of frequency 7 GHz, which, due to the modulation, corresponds to a spectral width $\sigma_f=1/(2\pi\sigma_t)\approx0.32$ GHz in the frequency domain. The areas of the cavities used in the experiment are $A=0.115 \textnormal{m}^2$ and 0.04$\textnormal{m}^2$ for the bowtie and cut-circle cavity, respectively. The corresponding typical ballistic flight times are about 1.3 ns and 0.7 ns. The Gaussian pulse was truncated to a total duration of about $6\sigma_t=3$ ns. To accumulate many runs of this basic process, we periodically broadcast $g(t)$ with a period $T=500$ ns $\gg \sigma_t$. The background noise level is about 2mV, and we set the input power to its maximum such that the sona signal $s(t)$ has a typical peak voltage of 150mV, much higher than the noise floor. $s(t)$ decays to the noise level within about 100 ns for the case of the bowtie cavity, because of ohmic loss in the upper and lower cavity plates and leakage through the ports. 

\par We carry out the synthetic sona calculation procedure for all orbits with orbit length less than $10\sqrt{A}$, where $A$ is the billiard area, a total of $1.2\times 10^5$ orbits for the bowtie billiard. We inject the time reversed synthetic sona (Fig.\ref{fig:fig2}(c)) into the microwave billiard to obtain the result at the target port shown in Fig.\ref{fig:fig2}(d).  Figure \ref{fig:fig2} (a) and (b) are the sona and the time-reversed reconstruction in the measured time-reversal scheme. The reconstruction signal shows a peak, which is the reconstructed Gaussian pulse, and symmetric sidelobes around the peak. Figure \ref{fig:fig2} (c) and (d) are the calculated synthetic sona (corresponding to orbits up to four meters long, or 15 ns) and its reconstruction at the target port in the microwave billiard. There is a significant peak in the reconstruction, but the sidelobes are now unbalanced. Nevertheless, this result demonstrates focusing at the target port in the experimental microwave billiard using a purely synthetic sona.

\subsection{Frequency Domain Experiment}
\label{sec:frequency_exp}
The time domain experiment setup described above takes at least 30 seconds to complete one time-reversal process with completely automated instrument control, and this imposes a constraint when we wish to systematically vary the carrier frequency of the input pulse. A sweep of carrier frequency from 1 GHz to 20 GHz takes hours, during which the cavity state may change due to temperature fluctuations or other time-dependant perturbations \cite{taddesea2009chaotic, bini1, bini2, bini3}. This problem can be addressed by switching to frequency domain measurements where the scattering parameter (S-matrix) of the system is measured only once and is then used to calculate the time-domain responses. The systems are linear and reciprocal, thus $S_{21}(\omega)=S_{12}(\omega)$ and $S(\omega)=G(\omega)S_{21}(\omega)$, $R(\omega)=S_{TR}(\omega)S_{21}(\omega)$ where $S(\omega)$, $G(\omega)$, $R(\omega)$ and $S_{TR}(\omega)$ are the frequency spectrum of $s(t)$, $g(t)$, $r(t)$ and $s(-t)$ respectively. The time domain sona and reconstruction signal can be obtained by calculating the inverse Fourier transform of $S(\omega)$ and $R(\omega)$. The output signal obtained in this way is the same as the one measured in the time domain experiment, albeit with much less noise. The signal-to-noise ratio of a $S_{21}(\omega)$ measurement is more than 30 dB while the time domain measurement (which measures $s(t)$ and $r(t)$ directly) has a constant background noise of about 2mV when the maximum peak voltage for $s(t)$ is 150mV. Later we will use this frequency domain version to explore the dependence of synthetic sona reconstructions on the center frequency of the Gaussian $g(t)$.

\section{Analysis}
\label{sec:analysis}
\subsection{Reconstruction Quality}
\label{sec:quality}
It has been shown by Derode, Tourin and Fink \cite{Lerosey1,fink05.2} that the reconstruction peak-to-noise ratio in a one-channel time reversal experiment scales as $\sqrt{\Delta f/\delta f}$ where $\Delta f$ is the effective bandwidth and $\delta f$ is the correlation frequency of the reverberated field. In our bowtie cavity experiment, for example, the bandwidth is about $2\sqrt{2\ln 2}\sigma_f=0.75$ GHz and the correlation frequency is governed by the Heisenberg time (the inverse of the mean spacing between eigenmodes), which is about 14.5 ns at 7 GHz. Hence the peak-to-noise ratio should be $\sqrt{0.75\times14.5}=3.3$, and the value observed in experiment is 2.78 in Fig \ref{fig:fig2}(b), comparable with expectations. 

Reference \cite{draeger1999one-exp} demonstrates that the reconstruction peak-to-noise ratio scales linearly with $\Delta T$, the length of sona used for time-reversal, when $\Delta T$ is small, and it saturates for larger $\Delta T$. In our case the loss is significant. The sona signal decays to noise level in about 100 ns while the recording time is 500 ns, thus we are already in the saturation region. Recording for a longer time only adds more background noise.

Here we discuss other factors that also influence the quality of a synthetic sona time-reversal reconstruction, for example, the propagation loss and port coupling mismatch. In order to examine the effect of these factors and compare reconstructions under different conditions, we first quantify the reconstruction quality using the following three metrics:
\begin{itemize}
\item The \textit{peak-to-peak voltage} $V_{pp}$ of the reconstructed Gaussian pulse. 
\item The \textit{focus ratio} is the average power of the reconstructed short Gaussian pulse divided by the average power of the entire reconstructed signal. The focus ratio measures how the reconstructed pulse stands out from the sidelobes and noise. For the case of perfect reconstruction, i.e., no sidelobes and noise, this quantity is equal to $(E/6\sigma_t)/(E/T)=T/6\sigma_t$ in the experiment described above, where $E$ is the energy of the signal. 
\item The \textit{transfer ratio} is the energy in the entire waveform that is received at the target port divided by the energy in the injected time-reversed sona signal. This metric quantifies how efficiently energy is being transfered from the transceiver port to the target port. 
\end{itemize}


\subsection{Effects of Loss and Mismatched Port Coupling}
\label{sec:loss}
Here we discuss two main factors that affect reconstruction quality: propagation loss and port coupling mismatch. Intuitively, a system with higher loss should lose more information during the transmission between the two ports, hence the reconstruction should be of lower quality. However, we have also observed that the time-reversal reconstruction in the superconducting cavity can be worse than that in a similar cavity in the normal state, mainly because of antenna coupling issues. Hence propagation loss and port coupling mismatch both affect reconstruction quality, and we now discuss them. 

\subsubsection{Effect of Loss on Reconstruction}
We find that to a good approximation the sona signal envelope decays exponentially in time as $e^{-t/\tau}$, where $\tau$ is the (assumed frequency-independent) sona amplitude decay time.  In particular, for the normal and superconducting cases $S_{\text{normal}}(t)\approx \mathcal{S}_{\text{normal}}(t)e^{-t/\tau_{\text{normal}}}$ and $S_{\text{sc}}(t)\approx \mathcal{S}_{\text{sc}}(t)e^{-t/\tau_{\text{sc}}}$, where $S(t)$ is the exponentially decaying sona signal and $\mathcal{S}(t)$ is the sona signal with an infinite decay time. Furthermore $\mathcal{S}_{\text{sc}}(t)$ and $\mathcal{S}_{\text{normal}}(t)$ are experimentally found to be approximately the same, with a cross-correlation coefficient of 0.92. Based on these results, in the case of the measured sona method, higher loss will result in a scaled down sona signal with faster decay rate, and, as we will next show, a scaled down reconstruction signal with smaller sidelobes; the reconstruction will thus have a smaller $V_{pp}$ and transfer ratio, but a higher focus ratio. 
\par Let $h(t)=h'(t)e^{-t/\tau}$ be the impulse response between the transceiver port and target port of the enclosure. In the case considered here the cavity is air-filled and waves travel nondispersively with velocity approximately $c=1/\sqrt{\mu_0 \epsilon_0}$; thus $h'(t)=\sum_j A_j\delta(t-t_j)$, $t\in (0,\infty)$, where $A_j$ and $t_j >0$ are the amplitude and the travel time along ray orbit $j$ connecting the transceiver port and the target port (with the convention $t_{j+1}>t_j$). The sona signal is $s(t)=g(t)*h(t),$ where ``$*$'' denotes convolution, and the reconstruction signal is $r(t)=s(-t)*h(t)=g(-t)*h(-t)*h(t),$ where $g(t)$ is the initial input signal. Thus,
$$ r(t)=\int g(-t')f(t-t') dt',$$ 
where $f(t)=h(-t)*h(t)=\sum_{j,j'}A_j A_{j'}e^{-(t_j+t_{j'})/\tau}\delta(t+t_j-t_{j'})$ is an array of Delta functions symmetric around $t=0$, which explains the balanced sidelobes around the peak in Fig.\ref{fig:fig2}(b). Thus 
$$r(t)=\sum_{j,j'}A_j A_{j'} g(t_{j'}-t_j-t)e^{-(t_j+t_{j'})/\tau},$$ 
and lower loss (larger $\tau$) leads to stronger reconstruction with larger $V_{pp}$, as expected. Also a larger $\tau$ leads to slower decay on both sides of the peak, making the focus ratio smaller.
\par
\begin{figure}
\includegraphics[width=0.45\textwidth]{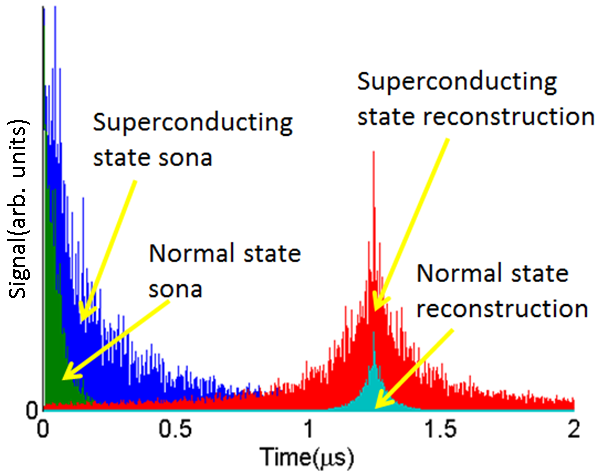}
\caption{\label{fig:fig5}Measured signals in cut-circle cavity in the measured sona time-reversal process. The sona signal in the cavity normal (green, shorter in time) and superconducting (blue, longer in time) states, and time-reversal reconstruction in the cavity normal (cyan, lower amplitude) and superconducting (red, higher amplitude) states. The sonas are generated by injection of a Gaussian pulse with $6\sigma_t=3$ns, modulating a 7 GHz carrier signal. Data is taken at 6.4 K (superconducting state) and room temparature (normal state).}
\end{figure}

 To better demonstrate the effect of loss on sonas and reconstructions, we compare the sona and the reconstruction measured in the cut-circle cavity in the normal and superconducting states, as shown in Fig.\ref{fig:fig5}. It is clear that the superconducting state sona, with $\tau_{\text{sc}}=153$ ns, has a much longer duration than the normal state sona, with $\tau_\text{normal}=50$ ns. The superconducting state reconstruction has a higher $V_{pp}$ but the focus ratio drops from 305 in the normal state to 158 in the superconducting state.  

\par For synthetic sona reconstruction, we replace the measured $h(t)$ from the time forward step with $h^{(s)}(t)=\sum_{j} A_j^{(s)}\delta(t-t_{j}^{(s)})$, where $t_{j}^{(s)}$ and $A_{j}^{(s)}$ are the calculated time delay and amplitude for the $j^{\text{th}}$ short orbit, respectively. Then, similar to $f(t)=h(-t)*h(t)$, we have $f^{(s)}(t)=h^{(s)}(-t)*h(t)=\sum_{j,j'}A_j^{(s)} A_{j'}e^{-(t_j^{(s)}+t_{j'})/\tau}\delta(t+t_j^{(s)}-t_{j'})$. Since the synthetic sona has a finite duration of $T_s$, $t_{j}^{(s)}$ must be in the range of $(0, T_s)$. So the prior-in-time sidelobe ($t < 0$) can only extend to $t=-T_s$. Hence, the sidelobes in the synthetic sona reconstruction are unbalanced, consistent with the result shown in Fig.\ref{fig:fig2} (d). If the synthetic sona duration is much shorter than the decay time $T_s\ll \tau$ then the left sidelobe, with length of $T_s$, will appear to be much shorter than the right sidelobe, which has decay time $\tau$, and thus leads to very poor reconstruction. Unbalanced sidelobes may also be seen in other situations, for example in one-bit time reversal \cite{derode1999onebit} where only the sign of the sona signal is recorded.
\par To summarize, higher loss (smaller $\tau$) results in a scaled down reconstruction signal with smaller sidelobes; the reconstruction will have a smaller $V_{pp}$ and transfer ratio, but a higher focus ratio. The synthetic sona reconstruction has unbalanced temporal sidelobes due to finite synthetic sona duration.

\subsubsection{Mismatched Port Coupling} 
\label{sec:mismatch}
Port coupling can be varied by using a different antenna or using different carrier frequencies for a given antenna. The former modifies the radiation impedance of the port entirely, and the latter uses the fact that radiation impedance is a function of frequency \cite{sameer1,zheng1}. Both effects lead to a different billiard transfer function $S_{21}(\omega)$, which is the ratio of the complex transmitted wave amplitude to the incident wave amplitude between the transceiver port (1) and the target port (2). Define the mean transmission $\mu\equiv\langle|S_{21}|^2\rangle_{\textnormal{avg}}$ averaged over a $6\sigma_f=2$ GHz frequency range surrounding the center frequency of the Gaussian pulse, and $\sigma_n\equiv\sigma(|S_{21}|^2)/\mu$ where $\sigma(|S_{21}|^2)$ is the standard deviation of $|S_{21}|^2$ in the same frequency range as $\mu$. $\mu$ and $\sigma_n$ measure the amplitude and fluctuations of the transmission spectrum $|S_{21}|^2$, respectively. We expect $\mu$ to have a linear relationship with $V_{pp}$ because $V_{pp}=r(0)$, $r(t)=g(-t)*h(-t)*h(t)=\int G(-\omega)e^{-i\omega t}|S_{21}|^2 d\omega$, thus setting $t=0$ leads to $V_{pp}\approx c\mu$ where $c$ is a voltage scaling factor.

\par
\begin{figure}
\includegraphics[width=0.45\textwidth]{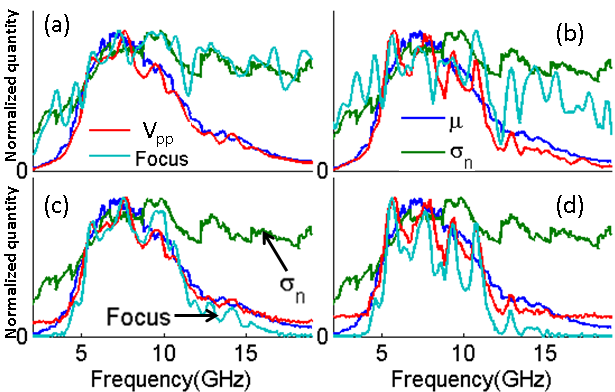}
\caption{\label{fig:fig3}Reconstruction quality ($V_{pp}$ and focus ratio) and $|S_{21}|^2$ statistics ($\mu$ and $\sigma_n$) measured in the 1/4-bowtie billiard as a function of carrier frequency: (a) (b) when no noise is added; (c) (d) when 2mV Gaussian random noise is added to the sonas and reconstructions. (a) (c) use the physically measured sona method, while (b) (d) use the synthetic sona method. All quantities are plotted normalized to their maximum values.}
\end{figure}

 Figures \ref{fig:fig3}(a) and (b) plot  $\mu$ and $\sigma_n$ as a function of pulse center frequency, together with the normalized $V_{pp}$, and the focus ratio for a series of measured reconstructions performed at the corresponding center frequencies. Figure \ref{fig:fig3} (a) is for the measured sona, and Fig.\ref{fig:fig3} (b) for the synthetic sona reconstructions. We find that $\mu$ and $\sigma_n$ predict the trend of $V_{pp}$ and focus ratio, respectively, in the physically measured sona method. The mean transmission $\mu$ has a peak around 7 GHz because the antenna is most efficient in that frequency range. For the synthetic sonas $\mu$ has a high correlation with $V_{pp}$, although $V_{pp}$ has stronger fluctuations compared to the case of a physically measured sona. 

\par Since this sweep over center frequency is done in the frequency domain as discussed in section \ref{sec:frequency_exp}, the calculated $s(t), r(t)$ is almost noise-free. To see the influence of noise on the reconstruction quality, we add Gaussian random noise with 2 mV standard deviation, the typical background noise in our time-domain experiment, to the synthetic sonas and the reconstruction signals calculated using the measured $S_{21}(\omega)$. Fig.\ref{fig:fig3} (c) and (d) show that when noise is added, $V_{pp}$ and focus ratio follow $\mu$ in both the measured sona and synthetic sona cases. This is because the average power in the reconstruction signal is mostly determined by the noise power, which is set to a constant, and the focus ratio is now proportional to average power in the reconstructed Gaussian pulse, which is proportional to $V_{pp}$. Hence $\mu$ becomes the only controlling factor in this case. 

\par In summary, knowledge of the mean value of transmission between the transceiver and target ports is an excellent predictor of reconstruction quality for both the physically measured sona and the synthetic sona methods. The higher the mean of $|S_{21}|^2$ in the given bandwidth of the pulse, the higher the quality of the reconstruction.

\subsection{Synthetic Sona Duration Constraint}
The synthetic sona duration is limited first by the computation cost and accumulation of error in the short orbit calculation. Since the number of orbits increases exponentially with orbit length, while the influence of each orbit decreases exponentially due to loss, it is more efficient to only calculate synthetic sonas with orbits within a length limit, depending on the computational budget. The ray-chaotic property of the billiard ensures that no ray is trapped inside the cavity without eventually reaching a port, but it also makes errors accumulate exponentially, at a rate determined by the largest Lyapunov exponent for the nonlinear map describing the ray trajectories. Hence the later part of the synthetic sona may contain more error than the earlier part.
\par
\begin{figure}
\includegraphics[width=0.5\textwidth]{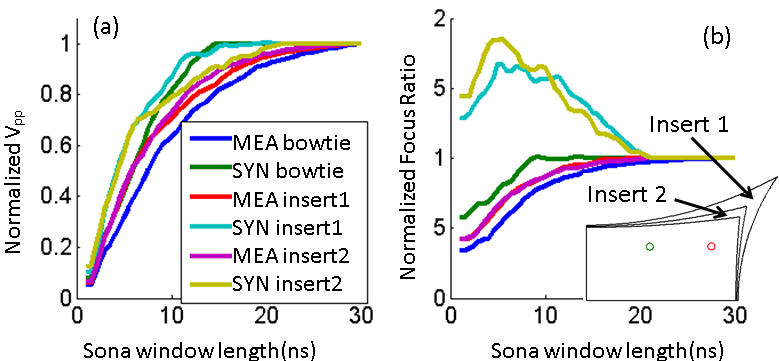}
\caption{\label{fig:fig4}Normalized reconstruction quality (normalized to its saturation value) for the physically measured and synthetic sona methods when a windowing function is applied to the sona before being time reversed, so that only the beginning part of the sona is used for the time backward step. The peak-to-peak voltage of the reconstruction is shown in (a) while the focus ratio is shown in (b). ``MEA" and ``SYN" refer to the physically measured and synthetic sona methods, respectively. ``insert1" and ``insert2" are two variations of the bowtie cavity geometry when inserts are added, as shown in the inset.}
\end{figure}

 To see the effects of accumulating errors, we create variations of the bowtie cavity by adding inserts to alter the scattering geometry of some of the walls. The differences between the geometry information of the actual inserts and the one assumed in the synthetic sona calculation are larger than that of the empty bowtie case. To determine the appropriate duration of the synthetic sona, we apply a windowing function to the full synthetic and measured sonas and plot the reconstruction quality (normalized to its saturation value) versus the sona duration in Fig.\ref{fig:fig4}. The windowing function has a 1.5 ns Gaussian-shaped rise and fall, to avoid introducing higher frequency components. For the measured sona method, both $V_{pp}$ and the focus ratio increase monotonically and eventually saturate when a longer sona duration is utilized. The saturation occurs when most of the sona signal with significant amplitude is used for time-reversal. The application of the windowing function is equivalent to changing $\Delta T$ as discussed in section \ref{sec:quality} and this behavior agrees with the findings in Ref \cite{draeger1999one-exp}. But for the synthetic sona method, the focus ratio is highest when the synthetic sona duration is around $4\sqrt{A}/c$(=4.5 ns) for the bowtie with inserts, where $c$ is the speed of light. This is because the shape of the inserts is known with less certainty than that of the empty bowtie, so the accumulation of error is more rapid. The later part of the synthetic sona contributes more to the sidelobes rather than to the reconstruction peak. The $V_{pp}$ of synthetic sona reconstruction also saturates eventually when all synthetic sona duration is utilized.
\par The synthetic sona duration is limited, but in order to have a good reconstruction the synthetic sona should be close to the $1/e$ amplitude decay time, $\tau$. We have shown in section \ref{sec:loss} that the duration of the earlier-in-time sidelobe (prior to the reconstruction peak) is determined by the synthetic sona duration, and the decay time of the later-in-time sidelobe (after the peak) is determined by $\tau$. If the synthetic sona is significantly shorter than $\tau$, then the reconstruction will have a small focus ratio with large sidelobes on the later side of the reconstruction, but very little sidelobe on the earlier side, causing it to look more like a sona signal rather than a reconstruction. For the bowtie billiard, the synthetic sona length is 15 ns which is close to $\tau=14$ ns, so it works well. But if we change to a less well-coupled antenna or decrease the propagation loss such that the decay time $\tau$ is much longer, the reconstruction quality drops significantly. This is confirmed with time-reversal experimental results from the superconducting cut-circle cavity which has a very long decay time $\tau_{\text{sc}}\approx153$ ns in the superconducting state and $\tau_{\text{normal}}\approx50$ ns in the normal state. The synthetic sona reconstruction in the superconducting state resembles a typical sona signal with a prominent sidelobe after the peak, while in the normal state it has balanced sidelobes, and thus better focus ratio. The focus ratio is 758 in the normal state with well-coupled antenna, 308 when changed to a less well-coupled antenna, and 123 when it is in the superconducting state with a less well-coupled antenna. For comparison, the focus ratio of an ideal reconstruction without sidelobe and noise in this experimental setup, as defined in section \ref{sec:quality}, is $T/6\sigma_t=$3333.

\section{Conclusion}
\label{sec:conclusion}
In this paper we have shown that focusing of electromagnetic waves at an arbitary location inside a ray-chaotic billiard can be achieved by using time-reversed synthetic sonas, calculated from the cavity geometry and location of the wave input and focusing points. The focusing quality is quantified and is influenced by cavity loss and port coupling. To achieve a high quality synthetic sona reconstruction with the optimal focus ratio, the billiard should be fairly lossy, and the synthetic sona duration should be close to the $1/e$ sona amplitude decay time, although it is limited by the computation cost and accumulation of error. In many practical applications, the systems are lossy (less reverberating), allowing for the synthetic sona to potentially work well. If the reconstruction amplitude or energy transfer is of more concern, then lower loss and better-coupled antennas (large mean transmission $\mu$) are required. 

\section{Acknowledgements}
We thank the group of A. Richter (Uni. Darmstadt) for graciously loaning the cut-circle billiard, H.J. Paik and M. V. Moody for use of the pulsed tube refrigerator, Martin Sieber for comments on improving the short orbit calculation algorithm, and the High-Performance Computing Cluster at UMCP for use of the Deepthought cluster. This work is funded by the ONR under Grants No. N00014130474 and N000141512134, and the Center for Nanophysics and Advanced Materials (CNAM). 

\bibliography{ref}

\end{document}